# FILE SYSTEM – A COMPONENT OF OPERATING SYSTEM

**Brijender Kahanwal*[1]**, Tejinder Pal Singh[2], Ruchira Bhargava[1], Girish Pal Singh[4]

[1]Department of Computer Science & Engineering, Shri Jagdishprasad Jhabarmal Tibrewala University, Jhunjhunu, Rajsthan imkahanwal@gmail.com

[2]Department of Applied Sciences, R. P. I. I. T., Bastara Karnal, Haryana

[3]Department of Computer Science and Information Technology, Maharaj Ganga Singh University, Bikaner, Rajasthan

## ARTICLE INFO

**Corresponding Author:**

Brijender Kahanwal Department of Computer Science & Engineering, Shri Jagdishprasad Jhabarmal Tibrewala University, Jhunjhunu, Rajsthan imkahanwal@gmail.com

**KeyWords:** File System, OS Component, Disk Controller, Buffer Cache, Hard Disk, Device Drivers.

## ABSTRACT

The file system provides the mechanism for online storage and access to file contents, including data and programs. This paper covers the high-level details of file systems, as well as related topics such as the disk cache, the file system interface to the kernel, and the user-level APIs that use the features of the file system. It will give you a thorough understanding of how a file system works in general. The main component of the operating system is the file system. It is used to create, manipulate, store, and retrieve data. At the highest level, a file system is a way to manage information on a secondary storage medium. There are so many layers under and above the file system. All the layers are to be fully described here. This paper will give the explanatory knowledge of the file system designers and the researchers in the area. The complete path from the user process to secondary storage device is to be mentioned. File system is the area where the researchers are doing lot of job and there is always a need to do more work. The work is going on for the efficient, secure, energy saving techniques for the file systems. As we know that the hardware is going to be fast in performance and low-priced day by day. The software is not built to comeback with the hardware technology. So there is a need to do research in this area to bridge the technology gap.

## INTRODUCTION

Data management is an important functionality provided by the operating system. File systems are tasked with volume of data management, including storing data on the disk or over the network, and naming. These are complex and very difficult to extend. The file system developers are hesitant to make major changes to them, because file system bugs may corrupt all the data on the secondary storage device. It is a difficult and time consuming job to develop a file system. A file system may be a general purpose, caching, cryptographic, compressing, or a replicated one which provides consistency control among the copies spread across the file systems.

The Figure 1 shows all the possible layers including hardware and the virtual views of spaces namely user space and the kernel spaces by the operating systems. This figure is further explained in detail in the following sections.

A file system is that part of an operating system that controls the storage and manipulation of files on media, such as disks. At first thought, that may seem like a rather straightforward task. However, to carry out this function, a file system supporting a multiuser operating system must perform a variety of difficult jobs.

A successful file system must do the following:

- ✓ Impose upon a blank medium a structure capable of representing highly organized data,
- ✓ Be able to multiplex the use of storage units among many concurrent processes,
- ✓ Contain internal synchronization for all processes,
- ✓ Enforce security by allowing data access only to those who have legitimate authority,
- ✓ Manage the sharing of individual files by multiple processes concurrently,
- ✓ Isolate faults stemming from imperfect physical media or improper access (This implies minimizing the potential for data loss due to hardware faults.), and
- ✓ Provide a standard set of interface routines to upper layers of the operating system and user programs and make them apply equally to devices with dissimilar interfaces.

In conclusion, a file system must do all these things with minimal impact on system performance.

## FILE SYSTEMS ELEMENTARY TERMINOLOGY

When we talk about file systems then there are many terms for referring to certain concepts, and so it is necessary to define how we will refer to the specific concepts that make up a file system. We list the terms from the ground up, each definition building on the previous.

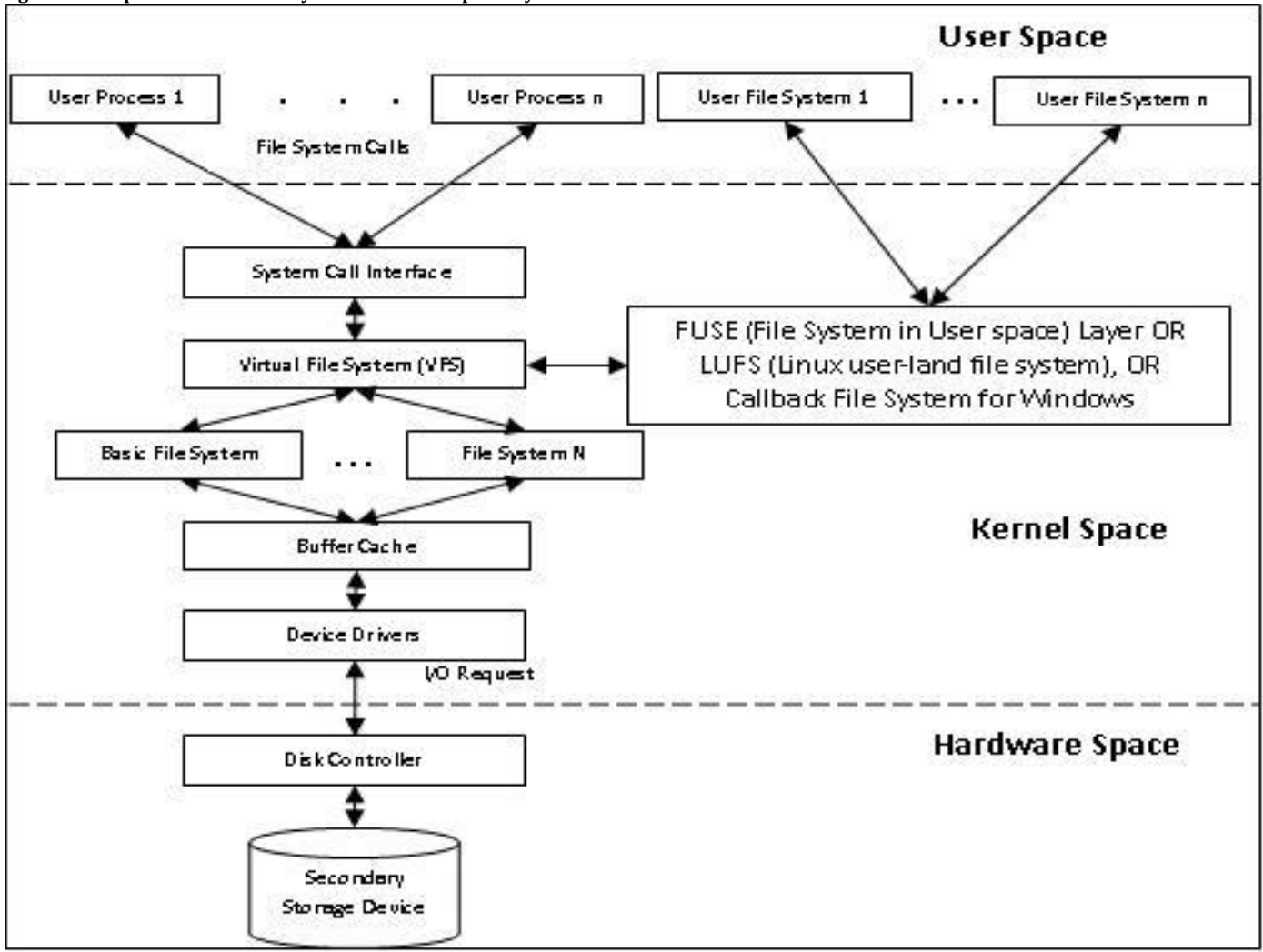

**Figure 1:** The positions of all file systems in the computer system.

✓ Disk: A permanent storage medium of a certain size. A disk also has a sector or block size, which is the minimum unit that the disk can read or write. The block size of most modern hard disks is 512 bytes.

✓ Block: The smallest unit writable by a disk or file system. Everything a file system does is composed of operations done on blocks. A file system block is always the same size as or larger (in integer multiples) than the disk block size.

✓ Partition: A subset of all the blocks on a disk. A disk can have several partitions.

✓ Volume: The name we give to a collection of blocks on some storage medium (i.e., a disk). That is, a volume may be all of the blocks on a single disk, some portion of the total number of blocks on a disk, or it may even span multiple disks and be all the blocks on several disks. The term "volume" is used to refer to a disk or partition that has been initialized with a file system.

✓ Superblock: The area of a volume where a file system stores its critical volume wide information. A superblock usually contains information such as how large a volume is, the name of a volume, and so on.

✓ Metadata: A general term referring to information that is about something but not directly part of it. For example, the size of a file is very important information about a file, but it is not part of the data in the file.

✓ Journaling: A method of insuring the correctness of file system metadata even in the presence of power failures or unexpected reboots.

✓ Inode: The place where a file system stores all the necessary metadata about a file. The inode also provides the connection to the contents of the file and any other data associated with the file. The term "inode" (which we will use in this book) is historical and originated in UNIX. An inode is also known as a file control block (FCB) or file record.

✓ Extent: A starting block number and a length of successive blocks on a disk. For example an extent might start at block 1000 and continue for 150 blocks. Extents are always contiguous. Extents are also known as block runs.

✓ Attribute: A name (as a text string) and value associated with the name. The value may have a defined type (string, integer, etc.), or it may just be arbitrary data.

**DEVICE DRIVER**

The device drivers have many definitions. These are written by the programmers. Essentially, it is simply a way to "abstract out" the details of peripheral hardware so the application programmer doesn't have to worry about them. In simple systems a driver may be nothing more than a library of functions linked directly to the application. In general purpose operating systems, device drivers are often independently loaded programs that communicate with applications through some OS-specific protocol. In multitasking systems, the driver should be capable of establishing multiple "channels" to the device originating from different application processes. In all cases though the driver is described in terms of an application programming interface (API) that defines the services the driver is expected to support.

The device driver paradigm takes on additional significance in a protected mode environment such as Linux. There are two reasons for this. First, User Space application code is normally not allowed to execute I/O instructions. This can only be done in Kernel Space at Privilege Level 0. So a set of device driver functions linked directly to the application simply would not work. The driver must execute in Kernel Space.

Actually, there are some loops we can jump through to allow I/O access from User Space but it is better to avoid them. The second problem is that User Space is swappable. This means that while an I/O operation is in process, the user's data buffer could get swapped out to disk. And when that page gets swapped back in, it will very likely be at a different physical address.

So data to be transferred to or from a peripheral device must reside in Kernel Space, which is not swappable. The driver then has to take care of transferring that data between Kernel Space and User Space.

## 1 DEVICE DRIVERS CATEGORIES

UNIX, and by extension Linux, divides the world of peripheral devices into three categories:
- ✓ Character
- ✓ Block
- ✓ Network

The principal distinction between character and block is that the latter, such as disks, are randomly accessible, that is, you can move back and forth within a stream of characters. Furthermore data on block devices is usually transferred in one or more blocks at a time and pre fetched and cached. With character devices the stream moves in one direction only. You can't for example go back and reread a character from a serial port. Block devices generally have a file system associated with them whereas character devices do not.

Network devices are different in that they handle "packets" of data for multiple protocol clients rather than a "stream" of data for a single client. Furthermore, data arrives at a network device asynchronously from sources outside the system. These differences necessitate a different interface between the kernel and the device driver.

## BUFFER CACHE

The purpose of the buffer cache is to increase system throughput by minimizing the time the CPU has to wait for disk I/O to complete. It does this by caching recently used data in the systems memory (read caching) and by delaying write operation (write caching).

The buffer cache is implemented as four queues of buffers, the locked queue, the clean queue, the dirty queue and the empty queue. A buffer contains information about the cached data such as the logical block number (The logical block number is an offset into the file, i. e. zero for the first block and one for the second etc.) and the physical disk block number and a pointer to the cached data. Each buffer can be used to cache between 512 byte (a sector) and 64 kilobyte worth of data but the file systems usually only reads and writes in complete file system blocks (defaults to 8 kilo byte for Fast File System).

The locked queue is used for important data and the buffers on this queue can't be recycled they are always in memory. The clean queue is for buffers that have invalid data, valid but unchanged data or data that has already been committed to the disk. The buffers in the clean queue are stored in Least recently Used (LRU) order which means that heavily accessed buffers will be stored longer in the cache and less used buffers will be recycled quickly. The buffers in the dirty queue are also stored in LRU-order and it contains data that needs to be written to disk. Write operations are delayed because a buffer that has been written to are likely to be written to again and by submitting all dirty buffers to the device driver simultaneous it can sort the write operation to minimize the seek time. The drawback of write caching is that in case of a system crash all unwritten data is lost. The risk of data loss is minimized by a special synchronization process that flushes all dirty blocks to disk every 30 seconds. Data in the dirty queue can also be flushed by a buffer cleaner process if the number of dirty pages exceeds 25 percent or the number of clean pages is less than 5 percent. The empty queue is for buffers that don't have any memory pages tied to them.

Table 1: The buffer cache interface

| System Call | Description |
|---|---|
| Bread | Read a block from a file. |
| Breadn | Read a block from a file and read n additional blocks asynchronously. |
| Baread | Read a block asynchronously. |
| Bwrite | Write a block to the file. |
| Bawrite | Write a block asynchronous to a file. |
| Bdwirte | Place the buffer on the dirty queue but don't start any I/O. |
| Brelse | Release a buffer back to the buffer cache. |

The buffer cache interface is presented in Figure 1. In general a file system reads data with bread or if the file is being read sequentially with breadn. This means that if a file is read in logical block number order when block number n is read it also starts to read block number n + 1 asynchronous so that when the user actually read block n + 1 it is already in memory.

File systems write data with the bdwrite function to increase performance but critical data is written with bwrite to maintain file system consistency in case of a crash. The bawrite is used by the buffer cleaner to start write operations on dirty buffers.

## HARD DISKS

Hard disk drives are classified as non-volatile, random access, digital, magnetic, data storage devices. Introduced by IBM in 1956, hard disk drives have decreased in cost and physical size over the years while dramatically increasing in capacity and speed.

A hard disk drive (HDD) is also named as a hard drive, or a hard disk, or disk drive. It is a device for storing and retrieving digital information, primarily computer data. It consists of one or more rigid (hard) rapidly rotating discs (platters) which are coated with magnetic material and with magnetic heads arranged to write data to the surfaces and read it from them.

In this subsection we introduce the terminology related to hard disks. In the Figure 5.1, important parts of the hard disk are shown. It consists of one or more circular aluminum platters, of which either or both surfaces are coated with a magnetic substance used for recording the data. For each surface, there is a read-write head that examines or alters the recorded data. The platters rotate on a common axis; typical rotation speed is 5400 or 7200 rotations per minute, although high-performance hard disks have higher speeds and older disks may have lower speeds. The heads move along the radius of the platters; this movement combined with the rotation of the platters allows the head to access all parts of the surfaces.

The processor (CPU) and the actual disk communicate through a disk controller that is discussed in the previous subsection. This relieves the rest of the computer from knowing how to use the drive, since the controllers for different types of disks can be made to use the same interface towards the rest of the computer. The controller may also do other things, such as caching, or automatic bad sector replacement.

The above is usually all one needs to understand about the hardware. There are also other things, such as the motor that rotates the platters and moves the heads, and the electronics that control the operation of the mechanical parts, but they are mostly not relevant for understanding the working principles of a hard disk.

The surfaces are usually divided into concentric rings, called tracks, and these in turn are divided into sectors. This division is used to specify locations on the hard disk and to allocate disk space to files. To find a given place on the hard disk, one might say ``surface 6, track 1, sector 9''. Usually the number of sectors is the same for all tracks, but some hard disks put more sectors in outer tracks (all sectors are of the same physical size, so more of them fit in the longer outer tracks). Typically, a sector will hold 512 bytes of data. The disk itself can't handle smaller amounts of data than one sector.

Each surface is divided into tracks (and sectors) in the same way. This means that when the head for one surface is on a track, the heads for the other surfaces are also on the corresponding tracks. All the corresponding tracks taken together are called a cylinder. It takes time to move the heads from one track (cylinder) to another, so by placing the data that is often accessed together (say, a file) so that it is within one cylinder, it is not necessary to move the heads to read all of it. This improves performance. It is not always possible to place files like this; files that are stored in several places on the disk are called fragmented.

The number of surfaces and the number of heads is the same thing. The cylinders, and sectors vary a lot; the specification of the number of each is called the geometry of a hard disk. The geometry is usually stored in a special, battery-powered memory location called the CMOS RAM, from where the operating system can fetch it during boot up or driver initialization.

Unfortunately, the Basic Input Output System (BIOS) has a design limitation, which makes it impossible to specify a track number that is larger than 1024 in the CMOS RAM, which is too little for a large hard disk. To overcome this, the hard disk controller lies about the geometry, and translates the addresses given by the computer into something that fits reality. For example, a hard disk might have 8 heads, 2048 tracks, and 35 sectors per track. Its controller could lie to the computer and claim that it has 16 heads, 1024 tracks, and 35 sectors per track, thus not exceeding the limit on tracks, and translates the address that the computer gives it by halving the head number, and doubling the track number.

The mathematics can be more complicated in reality, because the numbers are not as nice as here. This translation distorts the operating system's view of how the disk is organized, thus making it impractical to use the all-data-on-one-cylinder trick to boost performance.

Hard disk drives have been the main device for secondary storage of data in general purpose computers since the early 1960s. They have maintained this position because advances in their recording capacity, cost, reliability, and speed have kept pace with the requirements for secondary storage.

**Figure 2 A schematic figure of a hard disk.**

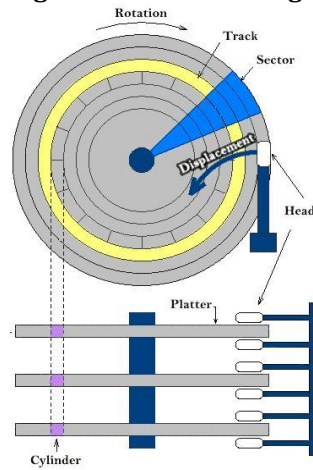

## DISK CONTROLLER

Storage is an important part of our computer system. In fact, most personal computers have one or more of the following storage devices:

- ✓ Floppy drive
- ✓ Hard drive
- ✓ CD-ROM drive

Usually, these devices connect to the computer through an Integrated Drive Electronics (IDE) interface. Essentially, an IDE interface is a standard way for a storage device to connect to a computer. IDE is actually not the true technical name for the interface standard. The original name, Advanced Technology Attachment (ATA), signified that the interface was initially developed for the IBM AT computer.

The disk controller is the circuit which enables the Central Processing Unit (CPU) to communicate with a hard disk, floppy disk or other kind of disk drive. It is a file that lets the operating system of our computer to communicate with the hard drives attached to the computer. The most common hard disk controllers are Intelligent Drive Electronics (IDE) and Small computer system interconnect (SCSI). IDE controllers are used in personal computers while SCSI is used in high end PCs, professional workstations, and network file servers.

The disk controller has its own CPU with Random Access Memory (RAM) buffer. It also has a Programmable Read Only Memory (PROM) that adds disk commands to Extended Color BASIC, a disk controller chip, and a little glue to make it all work.

It is the interface that enables the computer to read and write information to the hard disk drive. Today, hard disk drives have the controller built on to them, usually a circuit board that covers the bottom or on the back portion of the drive.

Early disk controllers were identified by their storage methods and data encoding. They were typically implemented on a separate controller card. Modified frequency modulation (MFM) controllers were the most common type in small computers, used for both floppy disk and hard disk drives. Run length limited (RLL) controllers used data compression to increase storage capacity by about 50%. Priam created a proprietary storage algorithm that could double the disk storage. Shugart Associates Systems Interface (SASI) was a predecessor to SCSI.

Modern disk controllers are integrated into the disk drive. For example, disks called "SCSI disks" have built-in SCSI controllers. In the past, before most SCSI controller functionality was implemented in a single chip, separate SCSI controllers interfaced disks to the SCSI bus.

The most common types of interfaces provided nowadays by disk controllers are Parallel Advanced Technology Attachment (PATA IDE) and Serial ATA for home use. High-end disks use SCSI, Fiber Channel or Serial Attached SCSI.

As can be seen in the above Figure 6.1, the bottom of the drive has a circuit board, which contains the hard disk controller of the laptop hard drive.

**Figure 3: The disk controller circuit board on the bottom part of the hard drive.**

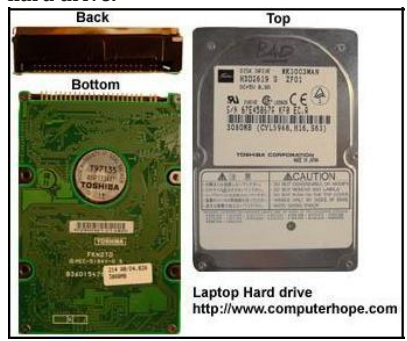

## CONCLUSION

This paper gives a light on the layer of file system in the computer system. All the upper layers and the lower layer are shown in it that gives a sound knowledge about the file system interaction between them. This review paper also tells about the view of file systems that may be the user file systems which has the interface kernel module like FUSE, LUFS, and Callback File System. These interface modules are intermediate between the VFS and the user space file systems.